\documentclass[twocolumn,showpacs,preprintnumbers,amsmath,amssymb]{revtex4}
\usepackage{graphicx}
\usepackage{dcolumn}
\usepackage{bm}
\usepackage{wasysym}
\usepackage{amsmath}
\usepackage{amssymb}

\newcommand{\apjs}{ApJS}

\newcommand{\mnras}{MNRAS}

\newcommand{\jcap}{JCAP}
\newcommand{\physrep}{Phys. Rep.}

\begin{document}

\title{Spectral analysis of the gamma-ray background near the dwarf Milky Way satellite Segue 1:
Improved limits on the cross section of neutralino dark matter annihilation}
\author{A. N. Baushev, S. Federici, and M. Pohl}
\affiliation{DESY, 15738 Zeuthen, Germany\\
 Institut f\"ur Physik und Astronomie, Universit\"at Potsdam, 14476
Potsdam-Golm, Germany}

\date{\today}

\begin{abstract}
The indirect detection of dark matter requires that dark matter annihilation products be discriminated from conventional astrophysical backgrounds. Here, we re-analyze GeV-band gamma-ray observations of the prominent Milky Way dwarf satellite galaxy Segue 1, for which the expected astrophysical background is minimal. We explicitly account for the angular extent of the conservatively expected gamma-ray signal and keep the uncertainty in the dark-matter profile external to the likelihood analysis of the gamma-ray data.
\end{abstract}


\maketitle

\section{Introduction}
The physical nature of dark matter is still mysterious. One of the most probable hypotheses is that
it is a mixture of equal quantities of weakly-interacting massive particles (WIMPs) and their
antiparticles, generated in the early Universe. The detection of their annihilation products could
be an attractive possibility to clarify the nature of these particles.

Dwarf satellites galaxies of the Milky Way are a class of objects that are very promising for
detecting dark-matter annihilation. Though the dark-matter density in these objects is
significantly lower than that at the Galactic Center, they typically contain no sources of cosmic
rays, and therefore almost any high-energy signal detected from the dwarf galaxies could be the
signal of dark-matter annihilation.

Tens of Milky-Way satellites are known to date. In our work we examine the dwarf galaxy Segue 1,
since it combines several advantages: it is the closest satellite to Earth, it is located at high
Galactic latitude (and, consequently, low gamma-ray background), and among the dwarf galaxies,
Segue 1 is expected to produce the strongest dark-matter signal
\cite{2009JCAP...06..014M,strigari,seguenew}. Due to it's proximity to the Sagittarius stream, the
nature of Segue 1 has been disputed: it was argued to be a disrupted star cluster originally
associated with the Sagittarius dSph \cite{2009MNRAS.398.1771N}. However, a kinematic study of a
larger member-star sample (66 stars compared to the previous 24-star sample) has recently confirmed
that Segue 1 is indeed a dwarf satellite galaxy \cite{seguenew}. Table~1 in \cite{fermi} shows that
the line-of-sight annihilation integral through the dark matter distribution of Segue 1 integrated
over the angular size of the source is the largest of all Milky Way dwarf satellites.

Other approaches used in the literature include performing a joint likelihood analysis of several
dwarf galaxies \cite{fermi} and a Bayesian technique \cite{2012arXiv1203.6731M}. We do not find a
joint analysis advantageous: actually, only a few objects can produce a flux comparable to Segue~1.
Therefore, the statistics can only be improved marginally, while the merging of several sources
with various properties, density-profile measurement errors, and backgrounds strongly increases the
ambiguities of the analysis.

Our consideration is based on the assumption that the WIMP in question is the lightest SUSY
particle, a neutralino. However, the only property of neutralinos we use is that they annihilate
mainly into a pair of heavy particles (heavy quarks, gauge bosons, tau leptons etc.). So our
conclusions are valid for any WIMPs for which the supposition applies.

We use data from the {\it Fermi} Large Area Telescope (LAT), since this telescope is currently the
most sensitive to the photons that can be generated by the annihilation of neutralinos with mass
roughly between 10~GeV and 1~TeV. Currently operating imaging Cherenkov telescopes are primarily
sensitive to gamma rays with energies $\gtrsim 200$~GeV and hence constrain very massive
dark-matter particles \cite{charbonnier}. {\it Fermi}-LAT data have already been used in several
articles \cite{fermi, bergstrom, strigari, 2011PhRvL.107x1303G}. However, all of them, except for
\cite{bergstrom}, considered Segue 1 as a point source. We perform an accurate spectral analysis
and explicitly account for the spatial extent of the source, using more data than \cite{bergstrom},
which allows us to improve the result. In \cite{fermi}, the Segue-1 dark-matter distribution is
modeled with a Navarro-Frenk-White profile \cite{1997ApJ...490..493N}, a centrally-concentrated
density distribution, thus leading to an enhanced gamma-ray signal.  We use a template for the
angular distribution based on the significantly more conservative Einasto profile with arbitrary
scale factor, diffuse and point-source backgrounds. The Einasto profile predicts a spatially more
extended signal, and the nominal annihilation flux is weaker than in \cite{fermi}. As we will show
in section \ref{sec:discussion}, the Einasto model predicts an annihilation signal that is only
marginally higher than the lowest limit for any density distribution compatible with the stellar
dynamics of Segue 1. We further investigate a possible position-dependent Sommerfeld enhancement
\cite{sommerfeld} of the annihilation rate. As both the total annihilation signal and the source
extent are affected by the choice of dark-matter profile and enhancement factors, the uncertainties
project onto the gamma-ray based upper limits in a non-trivial way.

\section{Model description}
In order to avoid ambiguity, we hereafter suppose that the dark-matter particle is identical to the
antiparticle. If the WIMP is a Dirac particle (which is not the case for neutralinos), then the
annihilation rate halves and all limits on the cross-sections would be weaker by a factor 2. On the
other hand, the main purpose of our work is to compare the limits obtained with the cross-section
estimate from dark-matter abundance $\langle \sigma v\rangle_a \simeq 3\cdot 10^{-26}$~cm$^3$
s$^{-1}$ \cite{bertone2005} (see below). This value is also increased by the same factor 2, if the
dark-matter particle is not a Majorana particle. Therefore, the ratio between the limit obtained
and $\langle\sigma v\rangle_a$ is not sensitive to the nature of the dark matter in this sense.

\begin{figure}[tb]
\centerline{\includegraphics[width=0.48\textwidth]{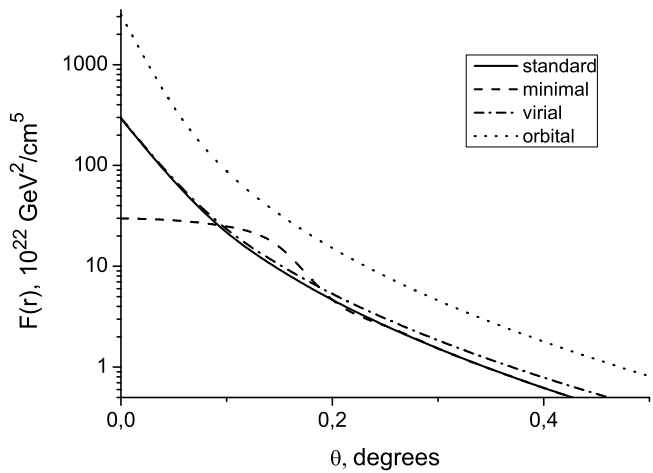}} \caption{The angular distribution of the signal for the four profiles used in this article. Here $\theta$ is the angular distance from the center of the source, $F(r)$ is defined by (\ref{12a2}) using $C=1$ and $\langle \sigma v\rangle=\langle \sigma v\rangle_c$.}\label{fig0}
\end{figure}

The annihilation signal (i.e., the number of photons $d\mathfrak N$ arriving at the local observer from a solid angle element $d\Omega$, per unit time interval $dt$, per an area $dA$, and an energy interval $dE$) in the general case is
\begin{equation}
I=\frac{d\mathfrak N}{dA d\Omega dE dt}= \dfrac{\langle\sigma\upsilon\rangle_c}{8\pi m^2_\chi}\cdot
F\cdot\frac{dN}{dE} \label{12a1}
\end{equation}
where $dN/dE$ is the spectrum of the photons produced by a single annihilation event, and
 \begin{equation}
F(r_p) =\int_{LOS} C \dfrac{\langle \sigma v\rangle}{\langle \sigma v\rangle_c} \rho^2 dl
 \label{12a2}
 \end{equation}
$F(r_p)$ is the line-of-sight integral of the multiplication of tree factors through the density
profile of the source at the projected radius $r_p$. $C=\langle\rho^2\rangle/\langle\rho\rangle^2$
is the so-called boost factor, which takes into account the  enhancement of the annihilation signal
by dark-matter substructures that may be present. The question of possible substructures in Segue 1
is extremely vague, and we will not consider it, accepting $C\equiv 1$. The factor ${\langle\sigma
v\rangle}/{\langle \sigma v\rangle_c}$ allows for a possible dependence of $\langle \sigma
v\rangle$ on particle speed, e.g. the Sommerfeld effect. $\langle \sigma v\rangle_c$ is a
characteristic value chosen such that for absent velocity dependence ${\langle\sigma
v\rangle}={\langle \sigma v\rangle_c}=$ constant.

With knowledge of the distance, $F(r_p)$ fully describes the angular distribution. We should mention that quantities such as $J(\Omega)\equiv L(\Omega)=\int_{\Omega(r)} F\left(r_p\left[\Omega\prime\right]\right) d\Omega\prime$, where $J$ or $L$ are considered as a function of the angle, are of more common use in the literature. Being defined so, they depend on the distance to the source, contrary to $F(r_p)$. We set the distance to Segue 1 to $23$~{kpc}.

The mass of Segue 1 within a certain radius is measured by observations of a few ($<100$) member
stars, but the density profile is assumed \cite[e.g.][]{2009JCAP...06..014M}, \cite{geha}. To
describe the signal profile we used four models. The first two assume that $\langle \sigma
v\rangle$ does not depend on the velocity.

The first model (hereafter the {\it standard} model)  is based on the Einasto density profile:
\begin{equation}
\rho=\rho_s \exp\left[-2 n\left\{\left( \frac{r}{r_s}\right)^\frac{1}{n} -1\right\}\right]
\label{12a3}
\end{equation}
with the parameters of Equation \ref{12a3} in accordance with \cite{magic} as
$\rho_s=1.1\cdot10^8$~$M_\odot$\,kpc$^{-3}$, $r_s=0.15$~{kpc}, and $n=3.3$. Further we assume for
the standard profile ${\langle\sigma v\rangle}={\langle \sigma v\rangle_c}$. It is worth mentioning
that the enclosed mass and the function $F(r_p)$ can be experimentally defined much better than the
individual parameters $\rho_s$, $r_s$, and $n$. Some authors \cite{bergstrom} use a completely
different set of parameters, which corresponds to a quite similar $F(r_p)$. We note that the total
integral $J_{tot}=\oint F d\Omega=10^{19.2}$~GeV$^2$\,sr\,cm$^{-5}$ in our case (and cgs units) is
smaller than the $J_{tot}=10^{19.6}$~GeV$^2$\,sr\,cm$^{-5}$ used in \cite{fermi}.

\begin{figure}[tb]
\centerline{\includegraphics[width=0.48\textwidth]{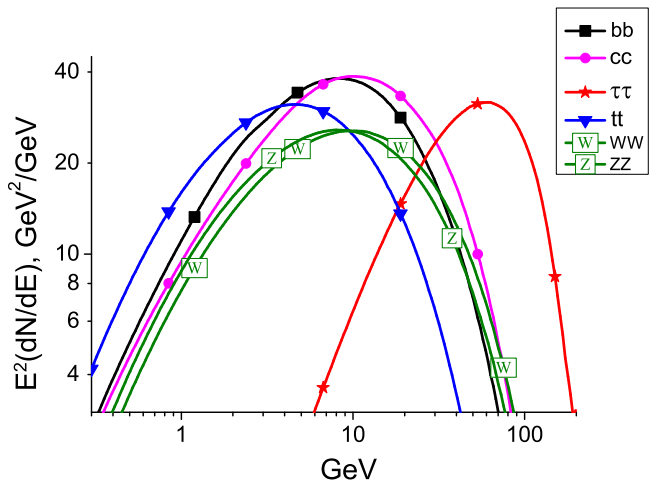}} \caption{The photon spectra generated by the annihilation of a pair of $200$-{GeV} neutralinos through various channels (the spectra are normalized as if the channel under consideration were the only channel of annihilation). The $b\bar b$, $c\bar c$, $\tau\bar \tau$, $t\bar t$, $W\bar W$, and $Z\bar Z$ are represented by the lines with squares, circles, stars, triangles, $W$, and $Z$, respectively.}\label{fig1}
\end{figure}

We should emphasize significant uncertainties of the Segue 1 density profile: almost all the stars
lie inside $10'$ ($\simeq 67$~pc) from the center \cite{seguenew}, which is significantly smaller
than $r_s$. In order to test how sensitive the annihilation signal is to the choice of profile, we
used a second ({\it minimal}) model. The assumptions of the {\it minimal} model are that the
density distribution beyond a radius of $67$~pc and the total mass inside this radius coincide with
those of the {\it standard} model (\ref{12a3}), but the dark matter is homogeneously distributed in
the inner $67$~pc. In this case the predicted signal is minimal among the profiles compatible with
observations. This conservative choice of density profile also renders the angular extent of the
annihilation signal larger than in the case of a more spiked density profile, for which a
point-source analysis may be suitable.

Among all the scenarios with $\langle \sigma v\rangle$ depending on $v$, we consider the Sommerfeld
effect \cite{sommerfeld}. If the dark-matter particles interact only via standard gauge bosons, the
effect arises only for heavy WIMPs, with masses in the TeV range, but if there are lighter mediator
particles, the enhancement appears even in the case of low-mass WIMPs \cite{sommerfeld2}. A similar
explanation has been invoked for the PAMELA positron excess \cite{pamela}; however, any discussion
of the PAMELA results or of the boosting necessary for the dark matter interpretation of the PAMELA
excess goes far beyond the scope of this article. We would like to emphasize only two essential
points. First, the Sommerfeld effect can take place only if there is some new relatively light
mediator field for the majority of the WIMP masses under consideration in this article (except for
the highest ones). Second, if the Sommerfeld effect takes place, $\langle \sigma v \rangle$ depends
on the collision velocity, and we should indicate the characteristic velocity when we obtain
experimental constrains of the cross-section.

To account for the Sommerfeld effect we used two models. Both of them are based on the Einasto
density profile, as the {\it standard} model. This allows us to find $\phi(r)$, the gravitational
potential of Segue 1. Roughly speaking, if the Sommerfeld effect occurs, then $\langle \sigma v
\rangle\sim 1/v$. The question is how $v$ depends on radius, $r$. We scale $\langle \sigma v
\rangle$ using as characteristic velocity the escape velocity from the center of Segue 1,
$v_c=2\,\sqrt{-\phi(0)}\simeq 60$~{km s$^{-1}$}:
\begin{equation}
\langle\sigma v\rangle=\frac{v_c}{v}\langle\,\sigma v\rangle_c\ .\label{12j1}
\end{equation}
For both the models taking into account the Sommerfeld effect the limits on the cross-section we
obtain are considered to mean the constraints on the quantity $\langle \sigma v \rangle_c$, the
$\langle \sigma v \rangle$ at the average collision velocity $v_c$. $\langle \sigma v \rangle$ at
the decoupling epoch could be much lower in this case.

Besides the absolute intensity, the Sommerfeld effect affects the angular profile: we should take
into account the velocity distribution of the particles as a function of $r$. The velocity
distribution properties are still not quite clear \cite{2009arXiv0911.3109B}. We consider two
possible hypotheses about the distribution of average collision speed $v$ inside Segue 1. We may
suppose that the speed is proportional to the escape velocity, or, for simplicity,
$v=2v_{esc}/\sqrt{2}=2\sqrt{-\phi(r)}$. A very similar situation takes place if radial motion of
the particles dominates \cite{2011MNRAS.417L..83B}. Hereafter we call this model {\it virial}.

In the opposite case, when the distribution function is supposed to be more or less isotropic, it
is reasonable to suppose $v=\sqrt{2} v_{orb}(r)$, where $v_{orb}(r)$ is the orbital speed at radius
$r$. This situation resembles the well-known isothermal dark-matter halo and should be typical for
the halo center \cite{profile}. Moreover, the experimentally observed \cite{seguenew} small
velocity dispersion of stars at the center of Segue 1, $\sim 3.7$~{km s$^{-1}$}, favors this
scenario. We name this model {\it orbital}.

The signal profile as described by each of the four models is shown in Figure~\ref{fig0}.

The error in the experimental determination of the multiplier $F(r_p)$ slightly depends on $r$ and can be estimated \cite{magic} as
\begin{equation}
\sigma\left(\log_{10} F\right)\simeq 0.6\ .
\label{12a4}
\end{equation}
If we assume the uncertainty in $\log_{10} F$ to be Gaussian, then the 2-$\sigma$ confidence
interval of $F(r_p)$ corresponds to an additional multiplier in equation (\ref{12a1}) extending
over the range $0.063-16$. It is obvious that this is the main source of uncertainty in the
determination of the upper limit on the cross section. However, $F(r_p)$ is an external factor in
the likelihood analysis of the {\it Fermi}-LAT data, provided the angular profile is realistic, and
therefore for a given dark-matter profile and signal boosting the limits on
$\langle\sigma\upsilon\rangle$ will to first order scale with the error factor in $F(r_p)$.

The angular size of the source is comparable to the angular resolution of {\it Fermi}-LAT above 10
GeV; for the {\it standard} model 68\% of the signal is expected within $0.17^\circ$ and 95\% of
the signal within $0.54^\circ$ of the source center. As we used a binned likelihood analysis, we
wanted to avoid a dependence of the model on the position in the map grid. For that purpose we
smoothed the profiles: Instead of $F(r_p)$ defined by (\ref{12a2}) we used $\tilde
F(r_p)\equiv\int_{\Delta\Omega}F(r_p)d\Omega /\Delta\Omega$, where $\Delta\Omega$ is $\sim
0.037^\circ\times 0.037^\circ$, which is slightly smaller than the pixel size and corresponds to an
area $15\text{pc}\times 15\text{pc}$ at the source. The smoothing kernel is about a factor 2.5
narrower than the angular resolution of {\it Fermi}-LAT above $10$~GeV and its particular choice
has no impact on the results of the likelihood fit.

The question of the photon spectrum produced by neutralino annihilation is, generally speaking, quite complex. Annihilation into a pair of particles (quarks, leptons, gauge bosons etc.) is the most likely process. The channels with light final states are suppressed by helicity, and the primary products of the process are heavy particles ($b$, $t$, $c$-quarks, $W$, $Z$-bosons, $\tau$ leptons). Let us denote the branching ratio of each individual channel by $k$ (for instance, $k_{\tau\tau}$ is the average fraction of annihilations resulting in a $\tau\bar\tau$ pair generation). The heavy primary particles are unstable and decay into stable products including photons (mainly from pion decays). We calculated the spectrum of photons generated by various annihilation channels with the DarkSUSY package \cite{darksusy}.

The real photon spectrum is a linear combination of the spectra corresponding to separate channels with coefficients $k_i$. Channel contributions $k_i$ (and, consequently, the spectrum) depend on the choice of SUSY model. To avoid a complete scan of the SUSY parameter space, we use the spectra of separate channels to obtain upper limits on the products $k_i \sigma$. For each neutralino mass we use two types of spectral models: the spectra corresponding to pure annihilation into $b\bar b$ and $\tau\bar \tau$, respectively. We reproduce in Fig.~\ref{fig1} the photon spectra generated by the annihilation of a pair of $200$-{GeV} neutralinos through various channels (the spectra are normalized as if the channel under consideration were the only channel of annihilation). As we can see, the spectra of all channels, except for $\tau\bar \tau$ and $t\bar t$, are very similar to that of the $b\bar b$ channel. Therefore, by placing a limit on $k_{bb}\sigma$, we actually provide a very good approximation of the upper bound on $(k_{bb}+k_{cc}+k_{WW}+k_{ZZ})\, \sigma$.

Note that, for a significant part of SUSY parameter space, $(k_{bb}+k_{cc}+k_{WW}+k_{ZZ})$ is close to $1$. The relative width of the $\tau\bar\tau$ channel is usually significantly smaller than $1$, but the spectrum is very hard and, consequently, easily distinguished, since the background rapidly decreases with energy. The $t\bar t$ spectrum is somewhat softer than that of $b\bar b$ decay, but we choose not to consider it as the third spectral model for a number of reasons. First of all, this channel is closed for light neutralinos because of the large mass of the $t$-quark. Second, $(k_{bb}+k_{cc}+k_{WW}+k_{ZZ})$ is typically at least comparable with $k_{tt}$ \cite{jungman}. Third, the spectrum of the $t\bar t$ channel does not differ very much from that of $b\bar b$.

We consider an array of neutralino masses that covers the expected mass range \cite{gorbrub1}
approximately uniformly in $\log E$: $40$, $50$, $63$, $80$, $100$, $126$, $159$, $199$, $251$,
$316$, $398$, $501$, $631$, $794$, $1000$, $1259$, $1585$, $1995$, $2512$, $3162$, $3981$, and
$5011$~{GeV}. In total we use 44 spectral models: a $b\bar b$ and a $\tau\bar \tau$ model for each
of 22 neutralino masses. The spatial models are, of course, independent on the neutralino mass.

\section{Observations of Segue 1 with {\it Fermi}-LAT}

We searched for evidence of gamma-ray emission from the dwarf spheroidal galaxy Segue 1 using data
from LAT, the principal instrument on board the {\it Fermi} spacecraft, a description of which is
given in \cite{2009ApJ...697.1071A}.

Data analysis was performed with the {\it Fermi}-LAT ScienceTools version 9.23.1, using the
post-launch P7SOURCE\_V6 instrument response functions. The cosmic-ray background was reduced by
choosing the P7SOURCE class which only includes events with high probability of being photons. We
also require a zenith-angle cut of 100$^\circ$ to minimize the spill-over of photons from the Earth
limb, excluding time intervals where any part of the region of interest (ROI) was beyond the
zenith-angle limit. In addition, data were not taken while the observatory was transiting the South
Atlantic Anomaly and excluded when the rocking angle exceeded 52$^\circ$. The data used in this
work came from observations in the period 2008 August 8 -- 2012 February 4.

A binned likelihood analysis is performed on a map with $0.05^{\circ}$ pixel size in gnomonic (TAN)
projection, measuring $10^{\circ}$ on a side and centered on $\alpha_{J2000}=151.767^{\circ}$ and
$\delta_{J2000}=16.0819^{\circ}$, the nominal position of Segue~1. We only used photons with
reconstructed energy greater than $2.4$~{GeV}, for which the 68\%-containment radius of the
point-spread function (PSF) is narrower than $\sim0.45^{\circ}$. The annihilation spectra are
always substantially harder than those of the background, and therefore a high energy threshold can
help to minimize background contamination.

\begin{figure}[tb]
\resizebox{\hsize}{!}{\includegraphics{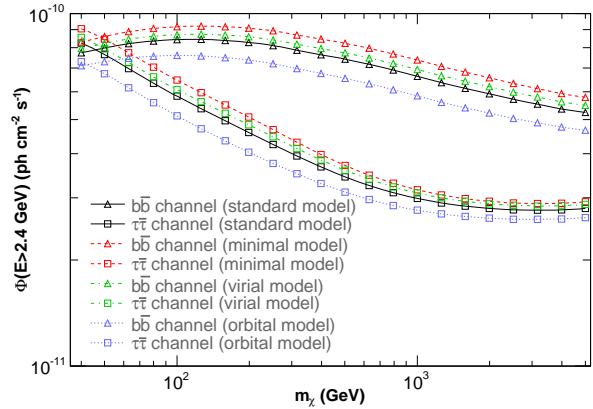}} \caption{95\%-C.L. integral flux upper limits for Segue 1. The curves with the same color represent the two channels 100\% $b\bar{b}$ (triangular markers) and 100\% $\tau\bar{\tau}$ (square markers) for a specific model. All of the four models described in the text are shown.
\label{fig:upplimit}}
\end{figure}

The resulting background model includes seven sources listed in the second LAT source catalog
\cite{2FGL}, and the LAT standard Galactic diffuse emission component \verb+gal_2yearp7v6_v0.fits+
along with the corresponding isotropic template \verb+iso_p7v6source.txt+ that accounts for
extragalactic emission and residual cosmic-ray contamination \footnote{Both are available from the
{\it Fermi} Science Support Center}. All seven sources lie outside our ROI and are modelled as
point sources with power-law energy spectra. Their spatial and spectral parameters were kept fixed
at the values given in the catalog, while we permitted the normalizations of the diffuse components
to freely vary.

Segue 1 is characterized by a spatial template of 2$^\circ$ diameter with a resolution of
$0.05^{\circ}$, following Equations \ref{12a2} and \ref{12a3}. Despite of the fact that the spatial
binning is slightly larger than the smoothing kernel applied to the Segue 1 profile, the results of
the likelihood analysis are not affected. Analyses with different binning scales show, in fact,
changes in the integral flux upper limits less than 1\%.

As described above, we use two different model spectra for the self-annihilation of neutralinos into quarks ($b\bar{b}$) and leptons ($\tau\bar{\tau}$) for each of the four model profiles ({\it standard}, {\it minimal}, {\it virial}, {\it orbital}). We tested 22 values of dark-matter particle mass in the range from 40 GeV to 5011 GeV.

\section{Analysis and results}\label{analysis}
We analyzed data between 2.4 GeV and 300 GeV. The low-energy limit was a priori chosen to optimize the constraints on the neutralino masses above 100 GeV. The normalization of the spatial template of Segue 1 is the relevant free parameter since we keep the position fixed. No significant gamma-ray signal was detected. We thus derive integral-flux upper limits over the energy range 2.4 GeV -- 300 GeV using the profile likelihood technique. Table \ref{tbl:1} provides 95\% confidence-level upper limits for dark-matter self-annihilation emission from Segue 1 ({\it standard} model only).

Figure \ref{fig:upplimit} shows the upper limits on the integrated flux as a function of the neutralino mass for the two annihilation channels for the four models. We note that the $b\bar{b}$ final state predicts on average a smaller photon energy than the $\tau\bar{\tau}$ final state (cf. Figure~\ref{fig1}). Therefore, the background contamination is more serious and, consequently, the flux upper limits are lower for the $\tau\bar{\tau}$ channel.

To be noted from Figure 1 is the effect of the shape of the annihilation profile, $F(r_p)$. The
30-\% difference between the best and worst upper limit reflect the deterioration in sensitivity
arising from source extent. The flux upper limits do not depend on the normalization of the
annihilation profile, the astrophysical factor $J$.

The impact of using P7SOURCE class events, instead of the older P6\_V3\_DIFFUSE, was also investigated. Differences of the two data sets slightly affect the results of the upper limits on the velocity-averaged self-annihilation cross section of neutralinos. The analysis performed with the Pass-6 data provides more constraining upper limits compared with the Pass-7 data, of around 10\% for the low-mass neutralino (i.e., m$_{\chi}<100$ GeV for $b\bar{b}$ and m$_{\chi}<501$ GeV for $\tau\bar{\tau}$ model) and around 20\% for those with higher mass. Note that this change is likely due to a statistical fluctuation. The templates for diffuse and isotropic emission are different for Pass~6 and Pass~7, and they dominate the likelihood fit. In the Pass-7 list of events between 2.4~GeV and 300~GeV reconstructed energy, we find 7 events within $0.5^\circ$ of the nominal position of Segue~1 (5 events in the Pass-6 event list). The entire region of interest comprises 920 events, or 9.2 per square-degree on average.

Comparing the standard profile with the point-source limit, we conclude that accounting for the
source extent in the likelihood analysis increases the upper limits by typically 20\%--25\%, but
only 10\%--15\% for $\tau\bar{\tau}$ models with large particle mass, for which most of the gamma
rays have energies $\gtrsim 100$~GeV.

\begin{table}
\begin{tabular}{|c|c|c|}
\hline
\footnotesize{$m_{\chi}$, GeV}& \footnotesize{Flux U.L., $b\bar{b}$} & \footnotesize{Flux U.L., $\tau\bar{\tau}$}\\
\hline
50   & 7.98 & 7.65 \\
100  & 8.43 & 5.83 \\
501  & 7.42 & 3.44 \\
1000 & 6.64 & 2.99 \\
5011 & 5.24 & 2.80 \\
\hline
\end{tabular}
\caption{Integral-flux upper limits above an energy threshold of 2.4 GeV for different WIMP masses ({\it standard} model). The confidence level is 95\% and the units $10^{-11}$ ph cm$^{-2}$ s$^{-1}$.\label{tbl:1}}
\end{table}

\section{Discussion}\label{sec:discussion}
\begin{figure}[tb]
\resizebox{\hsize}{!}{\includegraphics{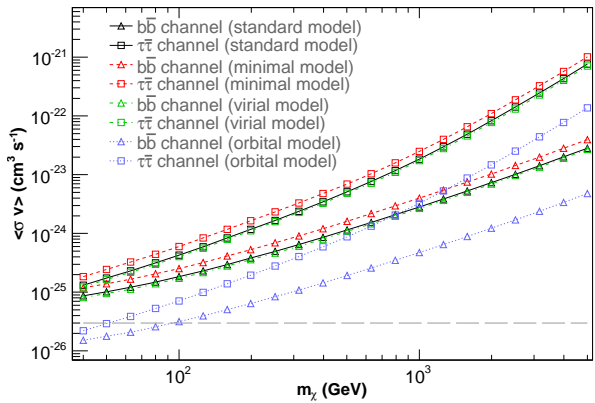}} \caption{95\%-CL upper limits on
$\left\langle\sigma v\right\rangle$ as a function of the neutralino mass for $k_{bb}=1$ (curves with triangular markers) and $k_{\tau\tau}=1$ (curves with square markers) for each of the four models. The horizontal dashed line shows the canonical value of the self-annihilation cross section for a thermal WIMP ($\left\langle\sigma v\right\rangle_a\simeq 3\times10^{-26}$ cm${^3}$ s$^{-1}$).\label{fig3}}
\end{figure}

Figure~\ref{fig3} represents the restrictions on $\langle\sigma\upsilon\rangle$ imposed by our
analysis, assuming $k_{bb}=1$ or $k_{\tau\tau}=1$. Note that the restrictions do not take into
account the uncertainty (Eq.~\ref{12a4}) in the dark-matter distribution. The systematic and
statistical uncertainties arising from modelling the stellar kinematics of Segue 1 are in fact the
dominant source of error, for which we do not know the probability distribution. In any case, it is
not a purely statistical uncertainty.

We can draw several conclusions. First of all, the limits for the {\it standard} and {\it minimal}
models differ only weakly, suggesting that the limits obtained using the Einasto profile are quite
conservative and not very sensitive to the shape of the density profile. The limits for the {\it
standard} and {\it virial} models almost coincide. This is a consequence of the fact that the
central region of the source is rather small, and its angular size is smaller than $r_s$. As a
result, $\phi$ (and so the average particle speed) changes only slightly inside the source.

The only model that excludes the relic abundance cross-section $\langle \sigma v\rangle_a \simeq
3\cdot10^{-26}$~cm$^3$ s$^{-1}$ (in the area of small neutralino masses) is the {\it orbital}. This
is due to the fact that $v_{orb}\to 0$ if the central density of the halo is finite. As a result,
the signal in the center is significantly boosted. Unfortunately, the limitation is significant
only for low-mass neutralinos and in the presence of the Sommerfeld enhancement. As we have already
discussed the Sommerfeld effect for low-mass particles can appear only if the WIMPs interact via
some low-mass mediator boson, discriminated from the known gauge bosons. So we can exclude only
rather exotic dark matter models.

A direct comparison of our results with those of \cite{fermi} is difficult, because they chose to
include in the likelihood analysis the substantial statistical uncertainty in the J-factor derived
from fitting a specific dark-matter profile to the stellar-kinematic data. Our limit on
$\langle\sigma\upsilon\rangle$ for the, e.g., $b\bar{b}$ channel with DM-mass 100~GeV is a factor 5
better than that of \cite{fermi}, and it is marginally lower than those quoted ibidem for Draco,
Ursa Major II, Ursa Minor, and Coma Berenices. As described in Section \ref{analysis}, using Pass-7
event selection and the extended source profile would increase the upper limit by $\sim 40\%$,
whereas the longer observing time should decrease it by a similar percentage, and thus the effects
should roughly balance each other.

The marginal improvement in the upper limits presented here compared with those in \cite{fermi}
quoted for Ursa Major II and Coma Berenices is therefore very reasonable. However, our nominal
$J$-factor for Segue 1 is a factor 2.5 lower than that used in \cite{fermi}, whereas their quoted
uncertainty in $J$ is a factor 3.4 which they include in the likelihood analysis. Again, one might
expect that accounting for the uncertainty in $J$ should roughly compensate the higher nominal $J$,
but our results suggest that we would need to set $\log J=18.5$ to match their upper limit, a value
that is 2$\,\sigma$ off their nominal $J$.

As we have already mentioned, the channels $({b\bar b}+{c\bar c}+{W\bar W}+{Z\bar Z})$ dominate for the majority of SUSY models, and the total spectrum differs little from that of ${b\bar b}$ annihilation. Therefore we may consider the limit given by the ${b\bar b}$ model to be with reasonable accuracy the limit on the total cross section.

But what constraints can the limits place on dark-matter models? First, all models for which the dark matter has never been in thermodynamic equilibrium should predict the annihilation cross section to be much lower than the limits shown in Figure~\ref{fig3}. If the dark matter was once in thermodynamic equilibrium, we can estimate the cross section at that epoch from the dark-matter abundance as $\langle \sigma v\rangle_a \simeq 3\cdot 10^{-26}$~cm$^3$ s$^{-1}$ for Majorana particles \cite{bertone2005}. As we can see, the upper limits never reach $\langle\sigma v\rangle_a$, except for the {\it orbital} model; for all other profiles they are at least a factor 4 higher.

On the other hand, the difference is not so large, and one might suppose that our results exclude
large cross-section enhancements, like strong Sommerfeld effect or high boost factor $C$.
Unfortunately, even this conclusion is, generally speaking, questionable. When dark matter was
formed, its average speed was tens of thousand of kilometers per second. Now it is of the order of
hundreds of kilometers per second \cite{2011MNRAS.417L..83B}. If the $s$-channel of annihilation
dominates (which is typically the case), $\langle \sigma v\rangle\simeq \mathrm{const}+\alpha v^2$.
The $p$-channel dominates more rarely, in which case $\langle \sigma \rangle\simeq
\mathrm{const}+\alpha v^2$ \cite{jungman}. We conclude that $\langle \sigma v\rangle$ for small
velocities, i.e. for the present epoch, can be similar to its value at freeze-out (in the first
case) or down to 100 times less (in the second case). In the latter instance, we cannot even
exclude a factor of a hundred enhancement. We may only disfavor a very large boosting ($>500$).

As we can see, the experimental results obtained do not significantly constrain dark-matter models
(except for the quite weak limitation on the boost-factor or Sommerfeld enhancement). A rational
question at this point is: What observations could improve the limit? First of all, a better
measurement of the dark-matter profile of Segue 1 would reduce the systematic uncertainties (cf.
Equation \ref{12a4}), but not necessarily the upper limits. However, even then the thermal
freeze-out cross-section $\langle \sigma v\rangle_a$ would not be challenged for any neutralino
mass, whereas we should reach at least this value to place any reasonable constraints on
dark-matter models. As we can see in Figure~\ref{fig3}, the limits on $\langle \sigma v\rangle$
must be improved by two orders of magnitude for particle masses around $1$~TeV.  At higher masses
the current generation of Cherenkov telescopes is more sensitive than {\it Fermi}-LAT
\citep{charbonnier}. For satellite-based detectors of GeV-band gamma rays this is hardly possible
because one would need to operate in the background-dominated mode, for which the sensitivity
increases only with the square root of the observing time. The isotropic, presumably extragalactic,
background alone provides an integrated photon flux (above 2.4 GeV) per 0.2$^\circ$-pixel that is
2.5\% of our $bb$-model upper limit. Improvements to the point-spread function of the gamma-ray
detector would not help on account of the intrinsic angular extent of the source. Observations of
gamma rays with energies $\sim 20$~GeV may still be conducted with moderate sky background,
although not in the photon-counting limit. Therefore, large Cherenkov telescopes such as CTA seem
to be a promising alternative. The main advantage of these instruments is their huge effective area
$\sim 10^6$~m$^2$ \cite{cta}, \cite{charbonnier} which could be enough to reach $\langle \sigma
v\rangle_a$ even for heavy WIMPs. The only significant weakness of Cherenkov detectors is their
relatively high energy threshold ($\sim 50$~GeV), i.e. they are fundamentally unable to detect the
annihilation of WIMPs of lower mass. On the other hand, for a large part of reasonable dark-matter
models (like mSUGRA or CMSSM) a low-mass WIMP will very soon be excluded (or detected) by the LHC.

\acknowledgments

Financial support by Bundesministerium f\"ur Bildung und Forschung through DESY-PT,
grant 05A11IPA,
is gratefully acknowledged. BMBF assumes no responsibility for the contents of this publication.

We acknowledge support by the Helmholtz Alliance for Astroparticle Physics HAP
funded by the Initiative and Networking Fund of the Helmholtz Association.

The \textit{Fermi} LAT Collaboration acknowledges generous ongoing support
from a number of agencies and institutes that have supported both the
development and the operation of the LAT as well as scientific data analysis.
These include the National Aeronautics and Space Administration and the
Department of Energy in the United States, the Commissariat \`a l'Energie Atomique
and the Centre National de la Recherche Scientifique / Institut National de Physique
Nucl\'eaire et de Physique des Particules in France, the Agenzia Spaziale Italiana
and the Istituto Nazionale di Fisica Nucleare in Italy, the Ministry of Education,
Culture, Sports, Science and Technology (MEXT), High Energy Accelerator Research
Organization (KEK) and Japan Aerospace Exploration Agency (JAXA) in Japan, and
the K.~A.~Wallenberg Foundation, the Swedish Research Council and the
Swedish National Space Board in Sweden.

Additional support for science analysis during the operations phase is gratefully
acknowledged from the Istituto Nazionale di Astrofisica in Italy and the Centre National d'\'Etudes Spatiales in France.

\end{document}